\documentclass[journal]{IEEEtran}

\usepackage{cite}      
\ifCLASSINFOpdf
  \usepackage[pdftex]{graphicx}
\else
  \usepackage[dvips]{graphicx}
\fi
\usepackage{amsmath}   
\usepackage{amssymb}   
\usepackage{amsfonts}  
\usepackage{bm}        
\usepackage{array}     
\usepackage{multirow}  
\usepackage{booktabs}  
\usepackage[ruled,vlined]{algorithm2e} 
\usepackage{url}       

\begin{document}

\title{A Switching Beamformer for Highly Non-Stationary Environments}

\author{Manan~Mittal,
        Ryan~M.~Corey,
        John~R.~Buck,
        and~Andrew~C.~Singer
\thanks{Manan Mittal is with Electrical and Computer Engineering, Stony Brook University, Stony Brook, NY 11794, USA (e-mail: manan.mittal@stonybrook.edu).}
\thanks{Ryan M. Corey is with Electrical and Computer Engineering, University of Illinois Chicago, Chicago, IL 60607, USA.}
\thanks{John R. Buck is with Electrical and Computer Engineering, University of Massachusetts Dartmouth, Dartmouth, MA 02747, USA.}
\thanks{Andrew C. Singer is with the College of Applied Science and Engineering, Stony Brook University, Stony Brook, NY 11794, USA.}
}

\markboth{ArXiv, June~2026}%
{Mittal \MakeLowercase{\textit{et al.}}: A Switching Beamformer for Highly Non-Stationary Environments}

\maketitle

\begin{abstract}
Adaptive beamforming is a cornerstone of array signal processing, yet its performance often collapses in the face of complex, rapidly changing interference. When interferers appear or move unpredictably, conventional estimators encounter a fundamental memory trade-off: short windows enable rapid tracking but suffer from high estimation variance, while long windows provide stable rejection but fail to adapt to shifts. This challenge is resolved by introducing the Universal Switching Beamformer (USB), which integrates competitive sequential prediction into the beamforming architecture. By employing a linear transition diagram, the USB implicitly maintains an exponentially large family of candidate covariance histories and dynamically re-weights them based on their cumulative output power. This mechanism allows the beamformer to automatically vary its effective memory length without explicit change detection or heuristic parameter tuning. A theoretical upper bound is proven on the regret relative to an omniscient oracle that selects the best piecewise-stationary covariance model in hindsight. Extensive simulations and experiments on the SwellEx-96 dataset demonstrate that the USB achieves the agility of short-window estimators and the precision of long-term integration, providing a principled solution for tracking highly non-stationary scenes.
\end{abstract}

\begin{IEEEkeywords}
Adaptive beamforming, array signal processing, non-stationary environments, sequential prediction, covariance matrices.
\end{IEEEkeywords}

\section{Introduction}
Beamforming constitutes a core methodology in array signal processing and underpins a broad spectrum of modern sensing and communication technologies. The objective is to process the measurements collected by multiple spatially distributed sensors in a manner that enhances a desired signal, while suppressing unwanted interference and noise. Classical formulations, and in particular the minimum power distortionless response (MPDR) beamformer \cite{capon1969high}, achieve this by constructing a spatial filter that preserves the response in the look direction while minimizing the total output power.

Adaptive beamforming's utility is fundamentally constrained by the dynamic nature of real-world environments. While the theory of optimal beamforming is mature \cite{van2002optimum, johnson1992array}, its practical implementation hinges on the accurate estimation of the environment's spatial covariance. In non-stationary environments, where interferers move, appear, or vanish abruptly, the covariance structure evolves on multiple timescales. A beamformer must therefore track these changes without sacrificing the statistical stability required for deep interference rejection.

Conventional adaptive beamformers attempt to track non-stationarity using sliding-window estimators or exponential forgetting. Both approaches impose an implicit memory length that governs how past observations influence the current covariance estimate. This choice dictates a rigid performance trade-off: a short window allows the algorithm to adapt quickly to environmental shifts but results in high estimation variance and noise sensitivity. Conversely, a long window provides low-variance estimates and stable nulls but responds slowly to transitions, often persisting in suppressing interferers long after they have moved. Because the ideal timescale of adaptation depends on unknown and variable environmental change rates, selecting a single fixed window length remains a persistent challenge in array signal processing.

To address this limitation, researchers have proposed heuristic schemes such as dominant subspace selection, adaptive forgetting factors and explicit change-point detection \cite{changepoint, cox1987robust, abraham1990beamforming, mittal2026time}. While these methods can improve tracking, they typically rely on pre-specified thresholds or assumptions about the magnitude of environmental shifts. When these assumptions fail, as in unpredictable or adversarial settings, the beamformer either overreacts to benign fluctuations or fails to track significant transitions, leading to performance degradation. It is argued that the problem of choosing an adaptation timescale is better framed through the lens of sequential decision theory, which evaluates performance without statistical assumptions on the data \cite{cesa2006prediction, merhav1998universal}. By comparing an algorithm's performance to the best beamformer chosen in hindsight, one can design adaptive systems that are robust against arbitrary patterns of interference variation. 

This competitive framework, which includes universal and competitive prediction schemes, ensures that a learner's cumulative error does not substantially exceed that of an oracle strategy, and develops bounds on the excess regret regardless of how the environment evolves \cite{kozat_singer}. These sequential decision theory algorithms provide rigorous performance bounds by comparing the learner to the best predictor in a reference class chosen in hindsight \cite{cesa2006prediction, buckblended}. 
 
The universal switching beamformer (USB) is presented to achieve this goal. At its core, a linear transition diagram encodes an exponentially large number of possible switching patterns among beamformers. Each path through this diagram represents a particular sequence of beamformer selections, including the possibility of switching at any time step. By maintaining a distribution over these paths and updating it sequentially as new data arrive, the algorithm simulates all possible switching schedules simultaneously. It implicitly evaluates the performance of each path, increasing the weight of paths that explain the data well. Through this sequential updating, the switching beamformer nearly matches the performance of the correct switching strategy chosen in hindsight, while the transition diagram's structure ensures computational efficiency.

This formulation links beamforming to universal sequential prediction, the USB acquires performance guarantees typically unavailable in classical beamforming theory \cite{reed1974rapid}. The algorithm satisfies a theoretical upper bound on regret relative to the best sequence of piecewise-constant covariance models selected in hindsight. This bound formally measures robustness against arbitrary patterns of interference variation, ensuring that the beamformer’s cumulative performance tightly tracks an oracle strategy possessing perfect knowledge of all future environmental changes. This analytical framework thus shifts the paradigm of adaptive beamforming: rather than relying on assumed statistical models, data-driven competitive performance is established against an adversarially varying environment. Beyond these theoretical guarantees, the formulation eliminates delicate parameter tuning and provides a unified mechanism that adapts seamlessly across stationary and nonstationary regimes. A complete implementation of the algorithm is provided to enable reproducibility and facilitate integration into existing signal processing systems.

\section{Signal Model}
Consider an array comprising $N_m$ sensors and $N_s$  sources. The sensors are assumed to be arbitrarily positioned within the environment, without any geometric constraints such as uniform spacing or planar arrangement. Despite this generality, centralized processing is assumed to be available, allowing all sensor measurements to be shared without restriction for joint beamforming.

Let $s_j$ denote the real-valued signal generated by the $j^{\rm th}$ source. As this signal propagates through the environment, it reaches sensor $m$ after passing through a sensor–source-specific transfer function $h_{m,j}$. Each sensor measurement also contains additive noise $v_m$, assumed to be spatially uncorrelated across sensors. Under these conditions, the observed signal at sensor $m$ and time $t$ is modeled as
\begin{equation}
x_m[t] = \sum_{j=1}^{N_s} (h_{m,j} \ast s_j)[t] + v_m[t],
\end{equation}
where $\ast$ denotes convolution.

Beamforming is performed by applying a linear filter to each sensor signal and summing the resulting outputs. Let $w_m$ denote the linear filter associated with sensor $m$. The beamformer output at time $t$ is therefore
\begin{equation}
z[t] = \sum_{m=1}^{N_m} (w_m \ast x_m)[t].
\end{equation}

For analytical convenience, the filter-and-sum structure is expressed in vector form. Each filter $w_m$ is assumed to possess a finite-length impulse response (FIR) of length $L$. The complete filter vector $\mathbf{w} \in \mathbb{R}^{N_m L}$ is constructed by stacking all filter coefficients in sensor-major order:
\[
\mathbf{w} =
\begin{bmatrix}
w_1[0] & w_1[1] & \dots & w_{N_m}[L{-}1]
\end{bmatrix}^\top
\]
Similarly, the most recent $L$ samples from each sensor are concatenated to form the observation vector $x[t] \in \mathbb{R}^{N_m L}$:
\begin{align*}
\mathbf{x}[t] = \big[ & x_1[t], \dots, x_1[t{-}L{+}1], \\
             & x_2[t], \dots, x_{N_m}[t{-}L{+}1] \big]
\end{align*}
Under this notation, the beamformer output reduces to the inner product
\begin{equation}
z[t] = \mathbf{w}[t]^T \mathbf{x}[t].
\end{equation}

The instantaneous second-order statistics of the array snapshot are captured by the instantaneous correlation matrix
$\mathbf{R}[t] = \mathbb{E}\big[\mathbf{x}[t] \mathbf{x}[t]^T\big] \in \mathbb{R}^{N_m L \times N_m L}$,
and all signals are assumed to be real-valued and zero-mean. Here $\mathbb{E}$ denotes expectation.

In many beamforming applications, partial knowledge of the desired source’s spatial signature is available. A common example is the set of weights that optimizes the white-noise gain, which coincides with the conventional beamformer for the known look direction. This beamformer can be viewed as the solution to the MPDR (Capon) problem under a diagonal-loading approximation, wherein only the diagonal of the covariance matrix is retained. Such formulations arise naturally in structured fields, for example, environments dominated by plane waves, where the array response can be synthesized directly from directional parameters.

In environments lacking clear geometric structure, or in reverberant settings where analytic steering vectors are unavailable, it is common to estimate the relative transfer function (RTF) between sensors. The RTF characterizes the relative gains and delays for the desired source and serves as an empirical counterpart to the steering vector, either obtained offline or estimated adaptively.

The subsequent development applies equally to both scenarios. Whether the steering vector is known a priori or derived from an RTF estimate, the corresponding steering weights are denoted by $\nu$ \cite{consolidated}. In the following section an upper bound is provided for the regret of the universal switching beamformer. Notably the produced bound is constructive such that the algorithm to implement the proposed method follows the proof and, by construction, satisfies the bound. 

\section{Universal Switching Beamformer}
In this section, a bound is derived on the regret for the USB algorithm.

\subsection{Deriving an Upper Bound}
\subsubsection{Leveraging probability assignment methods}
A performance bound is established for the universal switching beamformer (USB) relative to the best piecewise-stationary beamformer selected in hindsight. Consider a sequence of observations $\mathbf{\tilde{x}}^n = \{\mathbf{x}[1], \dots, \mathbf{x}[n]\}$ partitioned into $k+1$ stationary segments defined by the transition times $[t_0,t_1),[t_1,t_2),\dots,[t_k,t_{k+1})$ where segment $1$ is $[t_0,t_1)$ and segment $k+1$ is $[t_k,t_{k+1})$. Let $\mathcal{T}_{k,n}$ denote this specific segmentation path.

For a sequence of beamforming weight vectors $W = \{\mathbf{w}_1, \dots, \mathbf{w}_{k+1}\}$, where the weight vector $\mathbf{w}_i$ is fixed during the $i$-th segment $[t_{i-1}, t_i-1]$, the cumulative output power loss for this model (transition path and sequence of weight vectors) is given by
\begin{equation}
L_n(\mathbf{\tilde{x}}^n | W, \mathcal{T}_{k,n}) = \sum_{i=1}^{k+1} \sum_{j \in [t_{i-1}, t_i)} \left( \mathbf{w}_i^T \mathbf{x}[j] \right)^2.
\end{equation}
To facilitate the analysis within a sequential probability assignment framework, this loss is mapped to a ``pseudo-probability'' assigned to the data by the given model. This assignment relates the cumulative output power to an assigned pseudo-probability, or likelihood, defined as
\begin{equation}
P(\mathbf{\tilde{x}}^n | W, \mathcal{T}_{k,n}) \triangleq \exp \left\{ -\frac{1}{2\kappa} L_n(\mathbf{\tilde{x}}^n | W, \mathcal{T}_{k,n}) \right\},
\end{equation}
where $\kappa > 0$ is a user-defined scaling constant that controls the curvature of the loss function. This formulation allows us to treat the selection of the optimal beamformer weights and switching times as a maximum likelihood estimation problem.

For a fixed segmentation path $\mathcal{T}_{k,n}$, a competing oracle is defined as a sequence of locally optimal, fixed beamformers that maximize the likelihood of the observed data $\mathbf{\tilde{x}}^n$ within each stationary segment. By mapping the cumulative output power to a probability measure, the performance of this oracle can be expressed as the maximum possible likelihood assigned to the observation sequence. This probability is defined as
\begin{equation}
P^*(\mathbf{\tilde{x}}^n | \mathcal{T}_{k,n}) = \sup_W P(\mathbf{\tilde{x}}^n | W, \mathcal{T}_{k,n}).
\end{equation}
Substituting the maximum likelihood solution yields
\begin{equation}
P^*(\mathbf{\tilde{x}}^n | \mathcal{T}_{k,n}) = \exp \Bigg\{ -\frac{1}{2\kappa} \sum_{i=1}^{k+1} \sum_{t \in [t_{i-1}, t_i)} \left( (\mathbf{w}_i^*)^T \mathbf{x}[t] \right)^2 \Bigg\},
\end{equation}
where the optimal weight vector for the $i$-th segment, obtained by minimizing the squared error subject to the distortionless constraint $\boldsymbol{\nu}^T \mathbf{w}_i = 1$, is the standard Sample Matrix Inversion (SMI) Capon beamformer:
\begin{equation}
\mathbf{w}_i^* =
\frac{\left[ \hat{S}_{t_{i-1},\mathbf{\tilde{x}}^n}^{t_i - 1} \right]^{-1} \boldsymbol{\nu}}
{\boldsymbol{\nu}^T \left[ \hat{S}_{t_{i-1},\mathbf{\tilde{x}}^n}^{t_i - 1} \right]^{-1} \boldsymbol{\nu}}.
\end{equation}
Here, $\hat{S}_{t_{i-1},\mathbf{\tilde{x}}^n}^{t_i - 1}$ denotes the sample covariance matrix of the observations within the segment, defined as
\begin{equation}
\hat{S}_{t_{i-1},\mathbf{\tilde{x}}^n}^{t_i - 1} 
= \frac{\sum_{j \in [t_{i-1}, t_i)} \mathbf{x}[j] \mathbf{x}^T[j]}
{(t_i - 1) - t_{i-1}},
\end{equation}
and $\boldsymbol{\nu}$ represents the steering vector in the desired look direction.
Finally, the best global probability assignment is identified with $k$ transitions, by maximizing over all possible segmentation paths $\mathcal{T}_{k,n}$ with $k$ transitions:
\begin{align}
P^*(\mathbf{\tilde{x}}^n | \mathcal{T}_{k}^*) 
&= \sup_{\mathcal{T}_{k,n}} P^*(\mathbf{\tilde{x}}^n | \mathcal{T}_{k,n}) \\
&= \sup_{W,\mathcal{T}_{k,n}} P^*(\mathbf{\tilde{x}}^n | W, \mathcal{T}_{k,n}).
\end{align}
This expression corresponds to the probability assigned to the data by the best piecewise-stationary beamformer in the class of models exhibiting $k$ transitions.

\subsubsection{The strategy}
The objective is to determine a sequential beamforming algorithm capable of achieving a performance comparable to
\[
P^*(\mathbf{\tilde{x}}^n | W^*, \mathcal{T}_{k,n}^*)
\]
without prior knowledge of the number of transitions $k$ or the sequence length $n$.

This paper adopts a double mixture strategy. First, consider the probability assigned to the best beamformers for a fixed segmentation path $\mathcal{T}_{k,n}$. A Universal Dominant Mode Rejection (UDMR) \cite{buckblended, wage2018udmr} beamformer is employed for each stationary segment. The cumulative loss for the $i$-th segment is bounded by
\begin{multline}
\sum_{j \in [t_{i-1}, t_i)} \left( (\mathbf{w}_i^*)^T \mathbf{x}[j] \right)^2 
\leq \min_d \left( \sum_{j \in [t_{i-1}, t_i)} \left( \mathbf{w}_{d,i}^T \mathbf{x}[j] \right)^2 \right) \\
+ 2 A_x \, A_{w_i} \, \ln (D+1),
\end{multline}
where each $\mathbf{w}_{d,i}$ corresponds to a beamformer with dominant dimension $d$, and $D$ is the maximum rank.

The probability assignment is defined for a given path as
\begin{equation}
\tilde{P}(\mathbf{\tilde{x}}^n | \mathcal{T}_{k,n}) 
= \exp \left( -\frac{1}{2\kappa} \sum_{i=1}^{k+1} \sum_{j \in [t_{i-1}, t_i)} 
\left( (\mathbf{w}_i^*)^T[j] \mathbf{x}[j] \right)^2 \right).
\end{equation}
Taking the negative logarithm and applying the bound yields
\begin{align}
-2\kappa \ln \tilde{P}(\mathbf{\tilde{x}}^n | \mathcal{T}_{k,n})
&\leq \inf_W L_n(\mathbf{\tilde{x}}^n | W, \mathcal{T}_{k,n}) \nonumber \\
&\quad + 2 A_x \sum_{i=1}^{k+1} A_{w_i} \ln (D+1) \\
&\leq \inf_W \Big\{ L_n(\mathbf{\tilde{x}}^n | W, \mathcal{T}_{k,n}) \Big\} \nonumber \\
&\quad + 2 A_x \, A_w \, \ln(D+1) \cdot (k+1),
\end{align}
where $A_x$ bounds the squared norm of the input signal $\mathbf{x}[j]$, $A_{w_i}$ bounds the squared norm of the beamforming weight vector in segment $i$, and the global maximum weight bound is defined as $A_w = \max_i (A_{w_i})$.

Next, the universal probability distribution is defined $\tilde{P}_u(\mathbf{\tilde{x}}^n)$ as a mixture over all possible paths and segment counts:
\begin{equation}
\tilde{P}_u(\mathbf{\tilde{x}}^n) = \sum_{k=0}^{n-1} \sum_{\mathcal{T}_{k,n}} 
P(\mathcal{T}_{k,n}) \, \tilde{P}(\mathbf{\tilde{x}}^n | \mathcal{T}_{k,n}),
\end{equation}
subject to the constraints
\[
P(\mathcal{T}_{k,n}) \geq 0, 
\quad \sum_{k=0}^{n-1} \sum_{\mathcal{T}_{k,n}} P(\mathcal{T}_{k,n}) = 1.
\]
By construction, the universal probability satisfies
\[
\ln \tilde{P}_u(\mathbf{\tilde{x}}^n) 
\geq \ln P(\mathcal{T}_{k,n}) + \ln \tilde{P}(\mathbf{\tilde{x}}^n | \mathcal{T}_{k,n})
\]
for any specific transition path $\mathcal{T}_{k,n}$. Consequently, the universal loss is bounded by
\begin{multline}
-2\kappa \ln \tilde{P}_u(\mathbf{\tilde{x}}^n) 
\leq -2\kappa \ln P(\mathcal{T}_{k,n}) \\
+ \inf_W \Big\{ L_n(\mathbf{\tilde{x}}^n | W, \mathcal{T}_{k,n}) \Big\} 
+ 2 A_x A_w (k+1) \ln(D+1).
\end{multline}

\subsubsection{Assigning Path Probabilities}
The excess loss is now clearly a function of the probability assignment to the chosen path. To minimize this cost, the Krichevsky--Trofimov estimator is employed (KT) estimator \cite{krichevsky1981performance}, which assigns probabilities based on a mixture over Bernoulli parameters using the $Beta(0.5, 0.5)$ prior:
\begin{equation}
P_{KT}(e,b) \triangleq \int_0^1 \frac{1}{\pi \sqrt{\theta(1-\theta)}} (1-\theta)^e \, \theta^b \, d\theta.
\end{equation}
The KT estimator admits simple sequential updates, making it computationally efficient \cite{willems1996coding}:
\begin{align}
P_{KT}(e+1,b) &= \frac{e + \frac{1}{2}}{e+b+1} \, P_{KT}(e,b), \\
P_{KT}(e,b+1) &= \frac{b + \frac{1}{2}}{e+b+1} \, P_{KT}(e,b).
\end{align}
For any segmentation path $\mathcal{T}_{k,n}$ characterized by $k$ transitions (``ones'') and $n-k-1$ non-transitions (``zeros''), the path probability is defined as
\begin{equation}
P(\mathcal{T}_{k,n}) \triangleq 
\left( \prod_{i=0}^{k-1} P_{KT}(t_{i+1}-t_i-1,1) \right) 
P_{KT}(n-t_k,0).
\end{equation}
This probability assignment satisfies the redundancy bound
\begin{equation}
-2 A_x A_w \ln P(\mathcal{T}_{k,n}) \leq 2 A_x A_w \cdot \frac{3(k+1)}{2} \ln \frac{n}{k} + O(k).
\end{equation}

\subsubsection{The regret bound}
Combining this with the double mixture results derived earlier, an upper bound for the universal probability assignment is obtained:
\begin{multline}
-2 A_x A_w \ln \tilde{P}_u(\mathbf{\tilde{x}}^n) 
\leq \min_{W} \Big\{ L_n(\mathbf{\tilde{x}}^n | W, \mathcal{T}_{k,n}) \Big\} \\
+ 2 A_x A_w (k+1) \ln(D+1) \\
+ 2 A_x A_w \cdot \frac{3k+1}{2} \ln \frac{n}{k} + O(k).
\end{multline}
Equivalently, the universal regret is bounded in terms of the cumulative loss:
\begin{multline}
L_{n,u}(\mathbf{\tilde{x}}^n) \leq
\min_{W} \Big\{ L_n(\mathbf{\tilde{x}}^n | W, \mathcal{T}_{k,n}) \Big\} \\
+ 2 A_x A_w (k+1) \ln(D+1) \\
+ 2 A_x A_w \cdot \frac{3k+1}{2} \ln \frac{n}{k} + O(k).
\end{multline}

\subsection{Constructing a Sequential Algorithm}

To implement the universal beamformer efficiently, a mechanism to compute the conditional predictive density is required
\begin{equation}
\tilde{P}_u(\mathbf{x}[n] \mid \mathbf{\tilde{x}}^{n-1}) = \frac{\tilde{P}_u(\mathbf{\tilde{x}}^n)}{\tilde{P}_u(\mathbf{\tilde{x}}^{n-1})}
\end{equation}
in a sequential manner. By definition, the universal probability $\tilde{P}_u(\mathbf{\tilde{x}}^n)$ is a mixture over an exponentially growing number of segmentation paths $\mathcal{T}_{k,n}$. Direct evaluation of this sum is computationally intractable. However, the structure of the Krichevsky--Trofimov estimator allows this probability to be calculated recursively by organizing the paths into states based on their most recent transition.

\subsubsection{Bayesian Formulation}
The set of all possible switching paths is partitioned into $n$ disjoint sets. Let the state variable $s_n$ denote the time index of the most recent covariance reset (transition) prior to or at time $n$. That is, for any path $\mathcal{T}_{k,n}$ with transition times $\{t_1, \dots, t_k\}$, the state is defined as $s_n = t_k$. Under this definition, all paths that share the same last transition time $t_k = s$ are grouped into the same state $s_n = s$.

This grouping is sufficient because the optimal beamformer weight $\mathbf{w}^*_i$ for the current segment depends only on the data observed since the last transition. The history prior to $t_k$ influences the path probability weight but not the parameter estimate itself.

$P_n(s, \mathbf{\tilde{x}}^n)$ is defined as the total probability mass assigned to all paths ending in state $s$ at time $n$:
\begin{equation}
P_n(s, \mathbf{\tilde{x}}^n) = \sum_{\mathcal{T}': \, s_n = s} P(\mathcal{T}') \, \tilde{P}(\mathbf{\tilde{x}}^n \mid \mathcal{T}').
\end{equation}

\begin{figure}[t]
\centering
\includegraphics[width=0.8\linewidth]{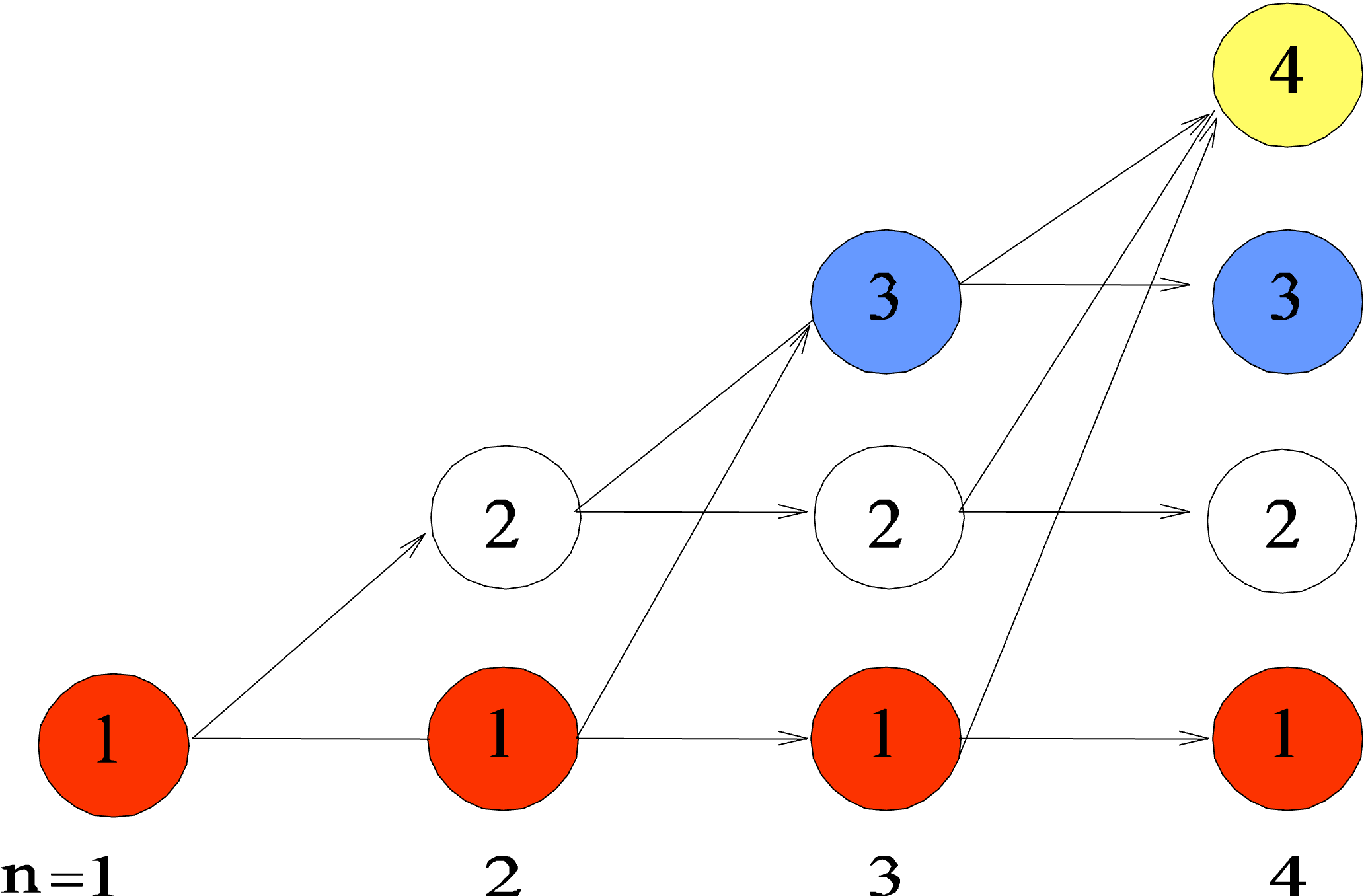} 
\caption{A graphical depiction of the linear transition diagram for $4$ time steps. At each time, the state may transition to a new state (upward transition) or remain in the same state (horizontal transition).}
\label{fig:ltd}
\end{figure}

\subsubsection{Double Mixture Approach}
The total universal probability is then the marginal sum over all possible current states:
\begin{equation}
\tilde{P}_u(\mathbf{\tilde{x}}^n) = \sum_{s=1}^n P_n(s, \mathbf{\tilde{x}}^n).
\end{equation}
This formulation reduces the complexity of the update from exponential in $n$ to linear in $n$. To compute $P_n(s, \mathbf{\tilde{x}}^n)$ recursively, a linear transition diagram is employed. Figure \ref{fig:ltd} shows the linear transition diagram for $n=4$. At each time step $n$, a process currently in state $s_{n-1}$ has only two possibilities:
\begin{enumerate}
    \item \textbf{Maintenance:} The process remains in the current state ($s_n = s_{n-1}$), implying no new changes occurred. The covariance estimate continues to update using the existing history.
    \item \textbf{Transition:} The process switches to a new state ($s_n = n$), implying a sudden change in the environment. This corresponds to resetting the covariance estimate.
\end{enumerate}
This structure allows the posterior probabilities of all candidate effective memory lengths to be propagated simultaneously.

The evolution of the state variable $s_n$ is governed by the transition probabilities derived from the Krichevsky--Trofimov estimator. At each time step $n$, a path currently in state $s_{n-1}$ can either persist in the same state (extending the current covariance history) or transition to a new state $s_n = n$ (resetting the covariance history). The probability of maintaining the current state $s_{n-1} = s$ is given by
\begin{equation}
P_T(s_n = s \mid s_{n-1} = s) = \frac{(n-1) - s + \frac{1}{2}}{(n-1) - s + 1},
\end{equation}
while the probability of switching to a new state is
\begin{equation}
P_T(s_n = n \mid s_{n-1} = s) = \frac{\frac{1}{2}}{(n-1) - s + 1}.
\end{equation}
These probabilities reflect the prior belief assigned to the continuation or termination of a stationary segment.

Next, the predictive likelihood of the observation $\mathbf{x}[n]$ is incorporated. For a path that does not switch, the likelihood is determined by the beamformer $\mathbf{w}_{s}^*$ trained on the history starting from $t=s$. The incremental update factor is
\begin{equation}
\label{eq:update_noswitch}
\rho(n, s) = \exp \left\{ -\frac{1}{2\kappa} \left( \mathbf{w}_{s}^{T}[n-1] \, \mathbf{x}[n] \right)^2 \right\},
\end{equation}
where $\mathbf{w}_{s}[n-1]$ denotes the UDMR beamformer estimated using data from the interval $[s, n-1]$.

Conversely, if a transition occurs ($s_n = n$), the history is reset. In the absence of a new covariance estimate, the beamformer reverts to the quiescent steering vector $\boldsymbol{\nu}$. The corresponding update factor is
\begin{equation}
\label{eq:update_switch}
\rho_0(n) = \exp \left\{ -\frac{1}{2\kappa} \left( \boldsymbol{\nu}^T \mathbf{x}[n] \right)^2 \right\}.
\end{equation}
Combining the transition priors and the predictive likelihoods allows the state probabilities to be updated the state probabilities recursively.

For any state $s < n$ (representing a preserved history), the probability mass comes exclusively from the same state at the previous time step:
\begin{equation}
\begin{split}
P_n(s, \mathbf{\tilde{x}}^n) &= P_{n-1}(s, \mathbf{\tilde{x}}^{n-1}) \, P_T(s \mid s) \\
&\quad \times \exp \left\{ -\frac{1}{2\kappa} \left( \mathbf{w}_{s}^{T}[n-1] \mathbf{x}[n] \right)^2 \right\}.
\end{split}
\end{equation}

For the new state $s = n$ (representing a reset), the probability mass is the aggregate of all paths that chose to switch at time $n$, regardless of their previous state. This forms the ``birth" node of the transition diagram:
\begin{equation}
\begin{split}
P_n(n, \mathbf{\tilde{x}}^n) &= \exp \left\{ -\frac{1}{2\kappa} \left( \boldsymbol{\nu}^{T} \mathbf{x}[n] \right)^2 \right\} \\
&\quad \times \sum_{j=1}^{n-1} P_{n-1}(j, \mathbf{\tilde{x}}^{n-1}) \, P_T(s_n = n \mid s_{n-1} = j).
\end{split}
\end{equation}

Consequently, the total universal probability at time $n$ can be expressed by marginalizing over the previous states:
\begin{equation}
\begin{split}
\tilde{P}_u(\mathbf{\tilde{x}}^n) &= \sum_{s_{n-1}=1}^{n-1} P_{n-1}(s_{n-1}, \mathbf{\tilde{x}}^{n-1}) \\
&\quad \times \Big[ P_T(s_n=s_{n-1} \mid s_{n-1}) \, P_n(\mathbf{x}[n] \mid \mathbf{\tilde{x}}^{n-1}, s_{n-1}) \\
&\quad + P_T(s_n=n \mid s_{n-1}) \, P_n(\mathbf{x}[n] \mid \mathbf{\tilde{x}}^{n-1}, s_n=n) \Big],
\end{split}
\end{equation}
where the state-dependent predictive likelihoods are defined as
\begin{equation}
\begin{aligned}
P_n(\mathbf{x}[n] \mid \mathbf{\tilde{x}}^{n-1}, s_{n-1}) &= \exp \left\{ -\frac{1}{2\kappa} \left( \mathbf{w}_{s_{n-1}}^{T}[n-1] \mathbf{x}[n] \right)^2 \right\}, \\
P_n(\mathbf{x}[n] \mid \mathbf{\tilde{x}}^{n-1}, n) &= \exp \left\{ -\frac{1}{2\kappa} \left( \boldsymbol{\nu}^{T} \mathbf{x}[n] \right)^2 \right\}.
\end{aligned}
\end{equation}

Hence, the sequential conditional probability is given by the ratio of the total probabilities:
\begin{align}
\tilde{P}_u&(\mathbf{x}[n] \mid \mathbf{\tilde{x}}^{n-1}) \nonumber \\
&= \frac{\tilde{P}_u(\mathbf{x}^{n})}{\tilde{P}_u(\mathbf{\tilde{x}}^{n-1})} \nonumber \\
&= \frac{\sum_{s_n=1}^{n} P_n(s_n, \mathbf{x}^{n})}{\sum_{s_{n-1}=1}^{n-1} P_{n-1}(s_{n-1}, \mathbf{\tilde{x}}^{n-1})} \nonumber \\
&= \frac{1}{\mathcal{Z}_{n-1}} \sum_{s_{n-1}=1}^{n-1} P_{n-1}(s_{n-1}, \mathbf{\tilde{x}}^{n-1}) \nonumber \\
&\quad \times \bigg\{ P_T(s_n=s_{n-1} \mid s_{n-1}) \nonumber \\
&\qquad\qquad \times P_n(\mathbf{x}[n] \mid \mathbf{\tilde{x}}^{n-1}, s_{n-1}) \nonumber \\
&\qquad + P_T(s_n=n \mid s_{n-1}) \nonumber \\
&\qquad\qquad \times P_n(\mathbf{x}[n] \mid \mathbf{\tilde{x}}^{n-1}, n) \bigg\},
\end{align}
where $\mathcal{Z}_{n-1} = \sum_{j=1}^{n-1} P_{n-1}(j, \mathbf{\tilde{x}}^{n-1})$ serves as the normalization factor from the previous time step.

To facilitate the analysis, the composite transition probability is defined as a density:
\begin{align}
P_{Tx}&(\mathbf{x}[n] \mid \mathbf{\tilde{x}}^{n-1}, s_{n-1}) \nonumber \\
&\triangleq P_T(s_n=s_{n-1} \mid s_{n-1}) \, P_n(\mathbf{x}[n] \mid \mathbf{\tilde{x}}^{n-1}, s_{n-1}) \nonumber \\
&\quad + P_T(s_n=n \mid s_{n-1}) \, P_n(\mathbf{x}[n] \mid \mathbf{\tilde{x}}^{n-1}, n).
\end{align}
Substituting this definition into the recursive update yields a mixture representation for the sequential conditional probability:
\begin{align}
\tilde{P}_u(\mathbf{x}[n] \mid \mathbf{\tilde{x}}^{n-1})
= \sum_{s_{n-1}=1}^{n-1} \mu_{n-1}(s_{n-1}) \, P_{Tx}(\mathbf{x}[n] \mid \mathbf{\tilde{x}}^{n-1}, s_{n-1}),
\end{align}
where $\mu_{n-1}(s_{n-1})$ represents the normalized posterior probability of the state $s_{n-1}$ given the past data:
\begin{equation}
\mu_{n-1}(s_{n-1}) = \frac{P_{n-1}(s_{n-1}, \mathbf{\tilde{x}}^{n-1})}{\sum_{j=1}^{n-1} P_{n-1}(j, \mathbf{\tilde{x}}^{n-1})}.
\end{equation}

The objective is to synthesize a single universal beamforming vector $\tilde{\mathbf{w}}_u[n]$ that satisfies the domination condition
\begin{equation}
\exp \left\{ -\frac{1}{2\kappa} \left( \tilde{\mathbf{w}}_u^{T}[n] \mathbf{x}[n] \right)^2 \right\}
\ge \tilde{P}_u(\mathbf{x}[n] \mid \mathbf{\tilde{x}}^{n-1}).
\end{equation}
Let the transformation function be defined as $f_t(z) \triangleq \exp\left(-\frac{z^2}{2\kappa}\right)$. This function is concave over the domain $z^2 \le \kappa = A_xA_w$. The conditional probability can be bounded applying Jensen's inequality:
\begin{align}
\tilde{P}_u&(\mathbf{x}[n] \mid \mathbf{\tilde{x}}^{n-1}) \nonumber \\
&\le f_t \Bigg( \sum_{s_{n-1}=1}^{n-1} \mu_{n-1}(s_{n-1}) \nonumber \\
&\quad \times \Big[ P_T(s_n=s_{n-1} \mid s_{n-1}) \, \mathbf{w}_{s_{n-1}}^{T}[n-1] \mathbf{x}[n] \nonumber \\
&\qquad + P_T(s_n=n \mid s_{n-1}) \, \boldsymbol{\nu}^{T} \mathbf{x}[n] \Big] \Bigg).
\end{align}

The argument of $f_t$ identifies the explicit form of the universal switching beamformer weight vector:
\begin{align}
\tilde{\mathbf{w}}_{u}[n] &= \sum_{s_{n-1}=1}^{n-1} \mu_{n-1}(s_{n-1}) \nonumber \\
&\quad \times \Big\{ P_T(s_n=s_{n-1} \mid s_{n-1}) \, \mathbf{w}_{s_{n-1}}[n-1] \nonumber \\
&\qquad + P_T(s_n=n \mid s_{n-1}) \, \boldsymbol{\nu} \Big\}.
\end{align}

Finally, the regret $R_W[n]$ is defined as the difference in cumulative loss between this universal algorithm and the best piecewise-stationary sequence selected in hindsight:
\begin{align}
R_W[n] &= \inf_{\substack{W_1,\dots,W_{k+1} \in \mathcal{W} \\ 1 \le t_1 < \dots < t_{k+1} = n+1}} \Bigg\{ \sum_{t=1}^{n} \left( \tilde{\mathbf{w}}_u^{T}[t] \mathbf{x}[t] \right)^2 \nonumber \\
&\quad - \sum_{i=1}^{k+1} \sum_{t \in [t_{i-1}, t_i)} \left( \mathbf{w}_i^{T}[t] \mathbf{x}[t] \right)^2 \Bigg\}.
\end{align}

Substituting the redundancy bounds inherent to the Krichevsky--Trofimov estimator and the universal parameter mixture, an explicit upper bound is obtained on the cumulative regret:
\begin{align}
R_W[n] \le &2 A_x A_w (k+1) \ln(D+1) + \\
&A_x A_w (3k+1) \ln\left(\frac{n}{k}\right) + O(k).
\end{align}

\subsection{Algorithmic Description}

This section presents an implementation of the Universal Switching Beamformer. To maintain computational tractability while adhering to the theoretical derivation, the algorithm utilizes a recursive update structure with a MPDR beamformer. The inverse correlation matrix for each state is updated via the Sherman-Morrison (Woodbury) identity, allowing for an efficient $\mathcal{O}(N_m^2)$ update per state rather than $\mathcal{O}(N_m^3)$. Additionally, a practical implementation would employ pruning, where the most likely states are maintained and updated. Table~\ref{tab:notation} summarizes the key symbols used in the algorithm.

\begin{table}[ht]
\caption{Algorithm Notation}
\label{tab:notation}
\vskip3pt
\centering
\begin{tabular}{cc}
\toprule
\textbf{Symbol} & \textbf{Description} \\
\midrule
$N_m$ & Number of sensors. \\
$\mathbf{x}[t]$ & Observation vector at time $t$. \\
$\boldsymbol{\nu}$ & Steering vector (Look direction). \\
$\tilde{z}_u[t]$ & Universal beamformer output at time $t$. \\
$\mathbf{P}_j[t]$ & Inverse correlation matrix (state $j$). \\
$\mathbf{w}_j[t]$ & Beamforming weight vector (state $j$). \\
$\mathbf{q}_j[t]$ & Unnormalized weight vector. \\
$\mu_j[t]$ & Posterior probability of state $j$. \\
$\tau_{j}[t]$ & Probability of maintaining state. \\
$\eta$ & Accumulated mass for new state. \\
$\kappa$ & Loss scaling parameter. \\
$\lambda$ & Diagonal loading factor. \\
\bottomrule
\end{tabular}
\end{table}

\begin{algorithm}[htbp]
\caption{Universal Switching Beamformer}
\label{alg:usb}
\SetKwInput{KwData}{Inputs}
\SetKwInput{KwResult}{Outputs}
\KwData{$\mathbf{x}[1{:}n]$ (observations), $\boldsymbol{\nu}$ (steering vector), $\lambda$ (loading)}
\KwResult{$\{\tilde{z}_u[t]\}$ (universal output)}

\textbf{Initialize:} \\
$\mathbf{P}_1 \gets \lambda^{-1} \mathbf{I}_{N_m}$, \quad $\mathbf{q}_1 \gets \lambda^{-1} \boldsymbol{\nu}$ \tcp*{Unnormalized weights}
$\mathbf{w}_1 \gets \boldsymbol{\nu} / (\boldsymbol{\nu}^T \boldsymbol{\nu})$, \quad $\mu_1 \gets 1$ \\
Define priors $P_{KT}(e, b)$ per Eq. (16).

\For{$t = 1$ \KwTo $n$}{
  $N_{states} \gets t$ \\
  
  \tcp{1. Beamforming and Universal Output}
  $z_{j}[t] \gets \mathbf{w}_j^T[t-1] \mathbf{x}[t]$ \tcp*{for $j=1..N_{states}$}
  $\tau_{j}[t] \gets \frac{(t-1) - t_{start}^{(j)} + 0.5}{(t-1) - t_{start}^{(j)} + 1}$ \tcp*{Transition prob}
  $\bar{\mu}_j \gets \mu_j / \sum_{i} \mu_i$ \\
  $\tilde{z}_u[t] \gets \sum_{j} \bar{\mu}_j \, \tau_{j}[t] \, z_{j}[t]$ \\

  \tcp{2. Measurement Update}
  Est.\ local noise power $\sigma_t^2$;\; $\eta \gets 0$ \\

  \For{$j = 1$ \KwTo $N_{states}$}{
    $\mathcal{L}_{j} \gets \exp\left( -\frac{z_{j}[t]^2}{2\sigma_t^2} \right)$ \tcp*{Compute likelihood}
    
    \tcp{Recursive Update (Woodbury)}
    $\mathbf{k}_j \gets \mathbf{P}_j \mathbf{x}[t] / (1 + \mathbf{x}[t]^T \mathbf{P}_j \mathbf{x}[t])$ \\
    $\mathbf{P}_j \gets \mathbf{P}_j - \mathbf{k}_j \mathbf{x}[t]^T \mathbf{P}_j$ \\
    $\mathbf{q}_j \gets \mathbf{q}_j - \mathbf{k}_j (\mathbf{x}[t]^T \mathbf{q}_j)$ \\
    $\mathbf{w}_j \gets \mathbf{q}_j / (\boldsymbol{\nu}^T \mathbf{q}_j)$ \\

    \tcp{Posterior Update}
    $\mu_j \gets \mu_j \cdot \mathcal{L}_{j} \cdot \tau_{j}[t]$ \\
    $\eta \gets \eta + \mu_j \cdot \mathcal{L}_{j} \cdot (1-\tau_{j}[t])$
  }

  \tcp{3. State Augmentation}
  $j_{new} \gets t+1$; \quad $\mu_{j_{new}} \gets \eta$ \\
  $\mathbf{P}_{j_{new}} \gets \lambda^{-1} \mathbf{I}$; \quad $\mathbf{w}_{j_{new}} \gets \boldsymbol{\nu} / (\boldsymbol{\nu}^T \boldsymbol{\nu})$ \\
  $\mathbf{q}_{j_{new}} \gets \lambda^{-1} \boldsymbol{\nu}$ \\
  Rescale all $\mu_j$ to prevent underflow.
}

\Return $\{\tilde{z}_u[t]\}_{t=1}^n$
\end{algorithm}

\section{Simulations}
To empirically validate the theoretical bounds established in previous sections and empirically benchmark the universal switching beamformer, the evaluation utilizes a comprehensive suite of numerical simulations. All compared beamformers use adaptive diagonal loading \cite{mittal2026adaptive}.

\subsection{Demonstrative Example}

\begin{figure}[t]
\centering
\includegraphics[width=\linewidth]{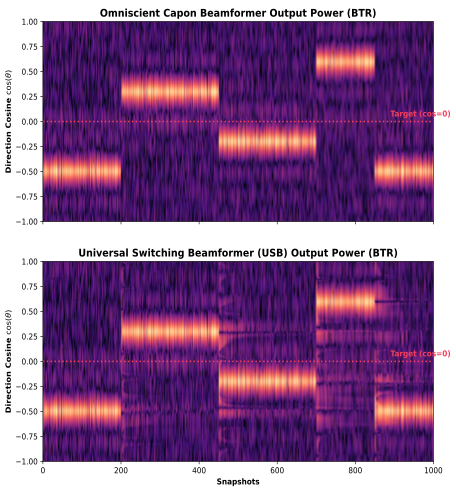} 
\caption{An image of the bearing time record for the USB and the Omniscient Capon beamformer for the demonstrative example of section IV-A.}
\label{fig:demonstrative_example}
\end{figure}

The fundamental behavior of the USB is first demonstrated in a simulated narrowband environment with piecewise-stationary interference. Sliding window algorithms struggle with sudden environmental changes, necessitating an evaluation of how the switching beamformer blends over these transitions. 

The simulation establishes an environment with a target signal arriving from broadside, corresponding to a direction cosine of $\cos(\theta) = 0$. Transient interferers shift spatial locations at discrete true switch times, specifically at snapshots $t \in \{200, 450, 700, 850\}$. The USB assesses the scene and updates its blend weights to best estimate the signal at the current snapshot. Figure \ref{fig:demonstrative_example} contrasts the estimated Bearing Time Record (BTR) of the USB against an omniscient Capon beamformer. 

\begin{figure}[t]
\centering
\includegraphics[width=\linewidth]{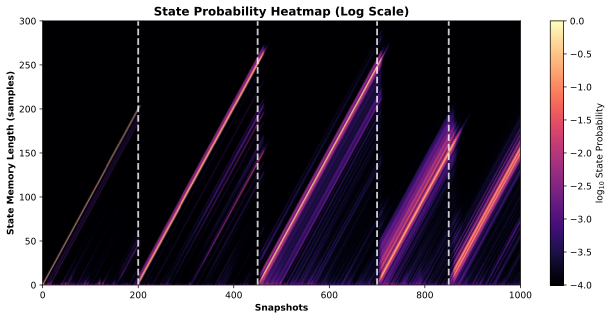} 
\caption{A heatmap of the state probabilities from the demonstrative example. As the sequence length grows, old states that no longer represent the environment, quickly have their probability set to 0.}
\label{fig:state_probabilities_heatmap}
\end{figure}

\begin{figure}[t]
\centering
\includegraphics[width=\linewidth]{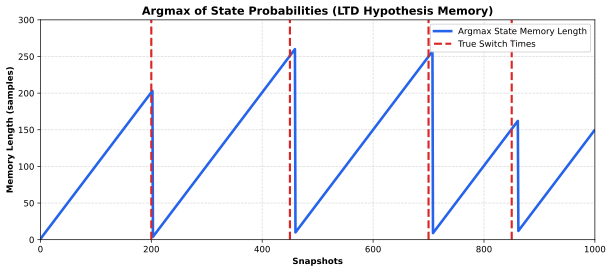} 
\caption{The argmax of the state probabilities over time. Note, the value of the function strongly corresponds with changes in the environment.}
\label{fig:argmax_state_probabilites}
\end{figure}

\begin{figure}[t]
\centering
\includegraphics[width=\linewidth]{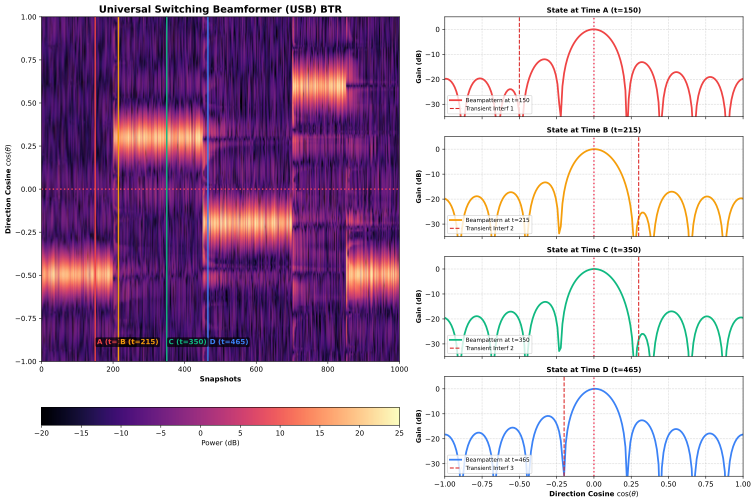} 
\caption{Demonstrative evaluation of the Universal Switching Beamformer (USB). The visualization depicts the spatial-temporal BTR and highlights the beampattern at denoted times. The USB quickly adapts to the time-varying environment and adaptively nulls active interferers.}
\label{fig:beampatterns}
\end{figure}

To visualize the USB’s internal dynamics, the evolution of the blend weights is examined across the linear transition diagram. Figure~\ref{fig:state_probabilities_heatmap} illustrates how the algorithm distributes weight over available states over time. Figure~\ref{fig:argmax_state_probabilites} tracks the $\text{arg max}$ of the PMF, confirming that the algorithm consistently identifies optimal states and tracks environmental switch points accurately. Finally, Figure \ref{fig:beampatterns} extracts instantaneous beampatterns at four analysis times: $t \in \{150, 215, 350, 465\}$.

\subsection{Adaptive Beamforming in Piecewise-Stationary Environments}

\begin{figure}[htbp]
\centering
\includegraphics[width=\linewidth]{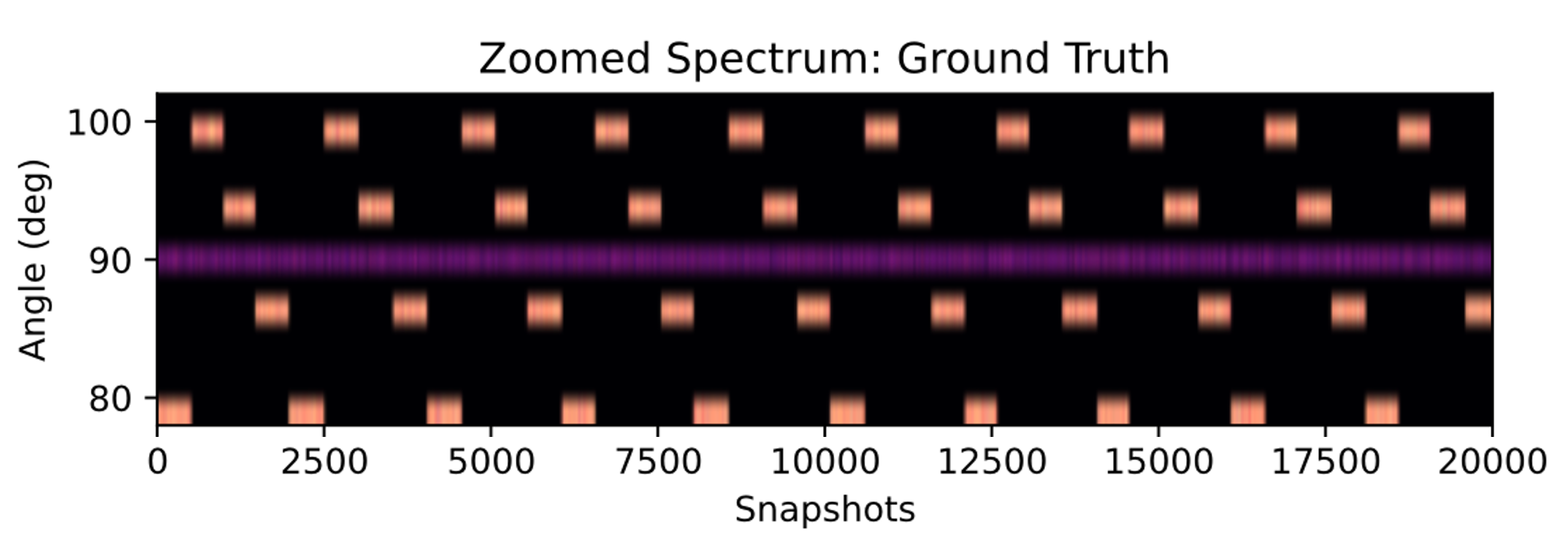}
\caption{The bearing-time record illustrating the piecewise-stationary in bearing interference environment, characterized by abrupt spatial transitions.}
\label{fig:pws_bearing_btr}
\end{figure}

\subsubsection{Piecewise Constant in Bearing}
The interference landscape is constructed from a predefined pool of four candidate arrival angles. A piecewise-stationary environment is enforced by activating exactly one interferer at any given moment. Figure~\ref{fig:pws_bearing_btr} visualizes the bearing-time trajectory of this scenario.

\begin{figure}[htbp]
\centering
\includegraphics[width=\linewidth]{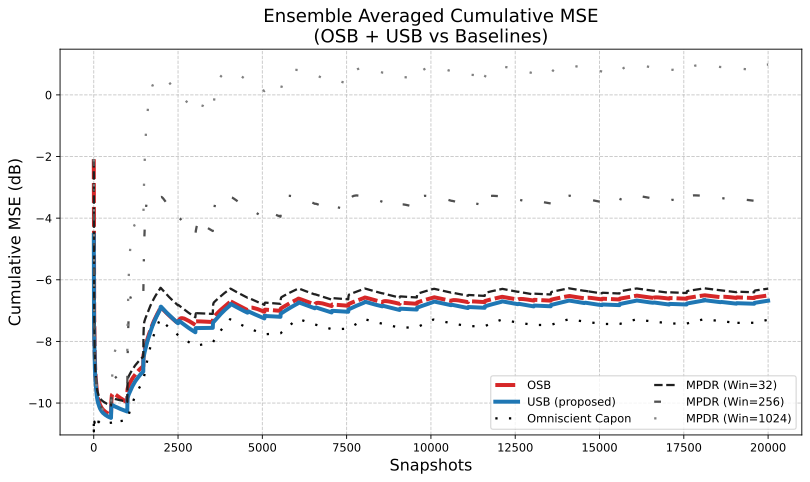}
\caption{Cumulative mean-squared error for estimating the desired signal in the piecewise-stationary spatial environment. The USB seamlessly navigates the bias-variance trade-off encountered by fixed-window methods.}
\label{fig:pws_bearing_mse}
\end{figure}

\begin{figure}[htbp]
\centering
\includegraphics[width=\linewidth]{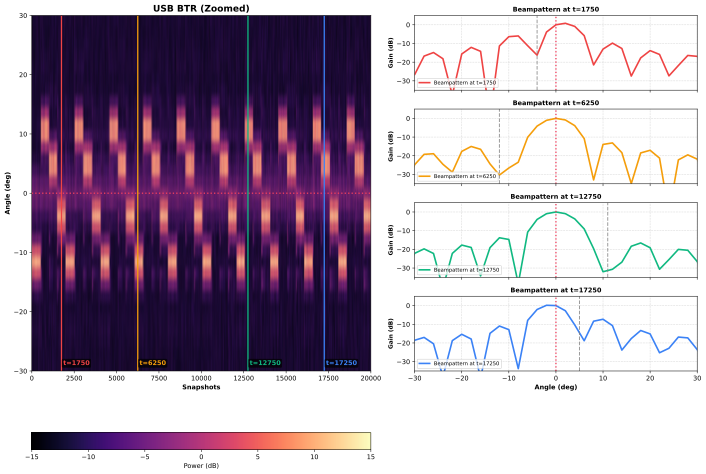}
\caption{Bearing time record and beampatterns at marked times for the universal switching beamformer for the example trial BTR for the piecewise stationary in bearing simulations.}
\label{fig:pws_bearing_beampatterns}
\end{figure}

The proposed USB is evaluated against a suite of conventional sliding-window MPDR beamformers, Omniscient Capon beamformer, the Conventional Beamformer (CBF), and the online segmented beamformer(OSB) \cite{mittal2026online, mittal2026time}. Figure~\ref{fig:pws_bearing_mse} presents the cumulative Mean Squared Error (MSE) averaged over $200$ Monte Carlo iterations. To further demonstrate this, Figure \ref{fig:pws_bearing_beampatterns} shows the bearing time record and the beampattern for the USB at marked times. 

\subsubsection{Adaptation to Irregular Temporal Scales}

\begin{figure}[htbp]
\centering
\includegraphics[width=\linewidth]{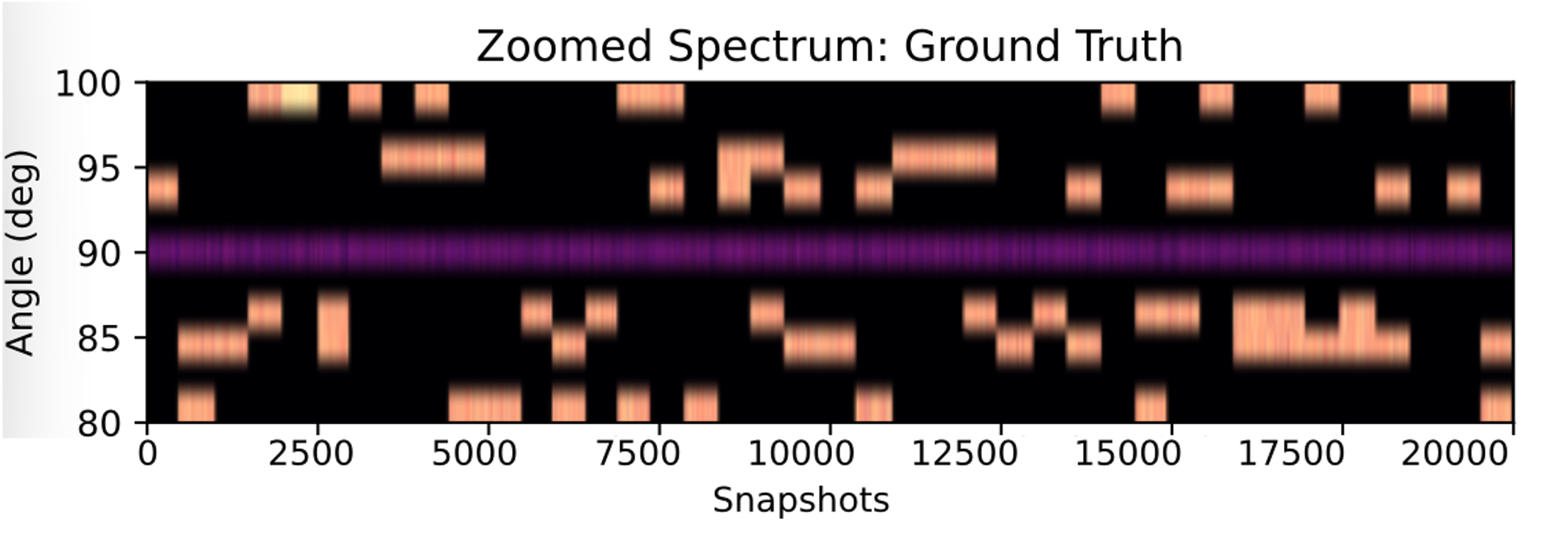}
\caption{The bearing-time record for the temporally variable scenario, highlighting the unpredictable duration of stationary interference blocks.}
\label{fig:pws_time_btr}
\end{figure}

A subsequent simulation investigates the algorithms' robustness against fluctuating timescales, defining a piecewise-constant time environment. The defining characteristic of this scenario is its high temporal variance: the duration of each dual-interferer block fluctuates wildly. The environment's ground truth temporal evolution for one trial is depicted in Figure~\ref{fig:pws_time_btr}.

\begin{figure}[htbp]
\centering
\includegraphics[width=\linewidth]{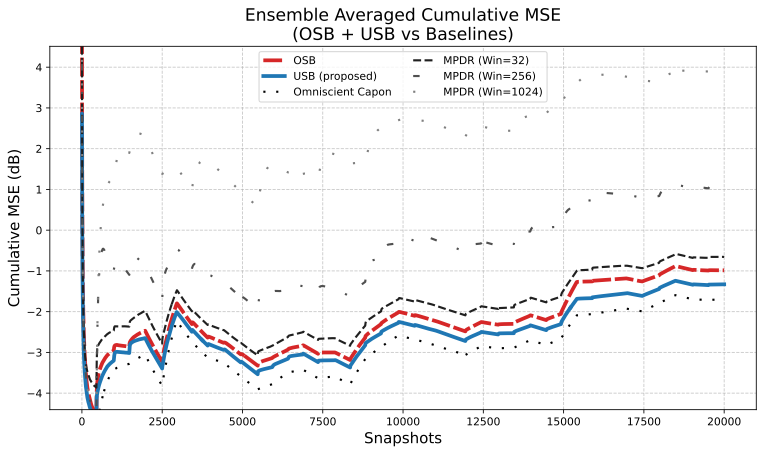}
\caption{Cumulative mean-squared error in the temporally fluctuating scenario. The USB natively adapts its effective integration time to match the irregular durations of the interference states.}
\label{fig:pws_time_mse}
\end{figure}

\begin{figure}[htbp]
\centering
\includegraphics[width=\linewidth]{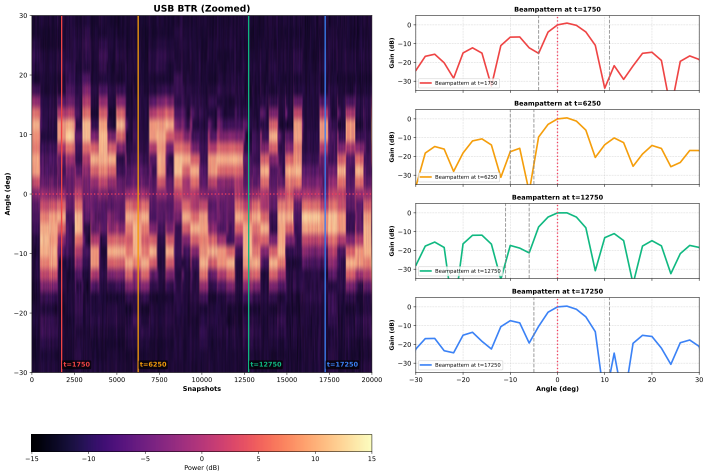}
\caption{Bearing time record and beampatterns at marked times for the universal switching beamformer for the example trial BTR shown for the piecewise stationary in time simulations.}
\label{fig:pws_time_beampatterns}
\end{figure}

The USB is benchmarked against the same fixed-window MPDR array and reference methods. Figure~\ref{fig:pws_time_mse} depicts the cumulative MSE evaluated across $200$ Monte Carlo trials. To inspect the behavior further Figure \ref{fig:pws_time_beampatterns} shows the bearing time record and the beampattern for the USB at the same times as the piecewise constant in bearing trials. 

\subsubsection{Tracking a Stochastic Birth-Death Interference Process}

\begin{figure}[htbp]
\centering
\includegraphics[width=\linewidth]{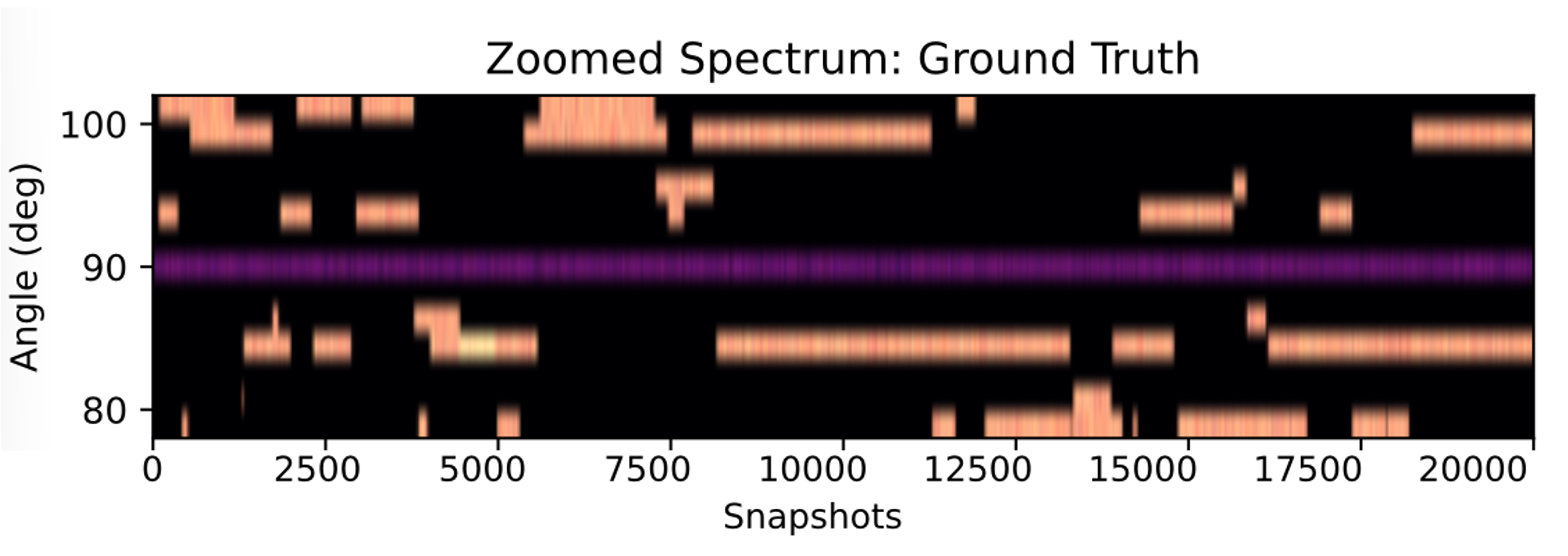}
\caption{Ground truth scanned response for one piecewise-stationary birth-death trial.}
\label{fig:birth_death_simulation_plots.png}
\end{figure}

To ascertain the real-time reliability of the proposed Universal Switching Beamformer in highly volatile settings, the interference field is modeled using a stochastic birth-death process. The activation and deactivation of these sources are governed by a Markovian birth-death framework. Figure~\ref{fig:birth_death_simulation_plots.png} illustrates the scanned response, demonstrating the USB's ability to maintain sharp spatial contrast alongside the omniscient baseline.

\begin{figure}[htbp]
\centering
\includegraphics[width=\linewidth]{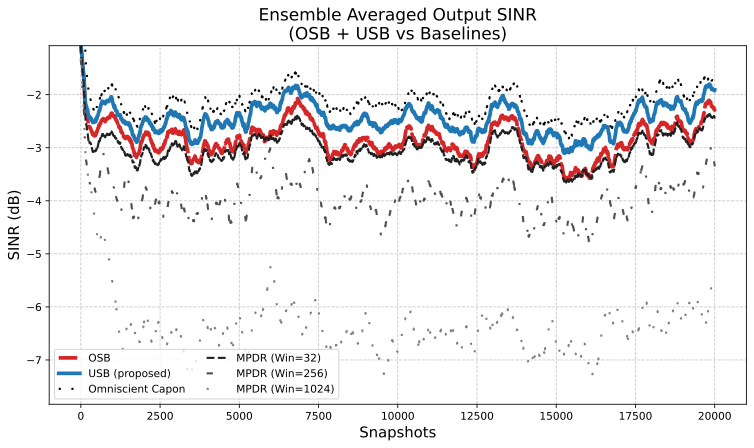}
\caption{Output signal-to-interference-plus-noise ratio (SINR) for the birth-death Monte Carlo trials. The USB swiftly redirects probability mass to reset states during new interference onsets, rapidly recovering optimal SINR.}
\label{fig:birth_death_sinr_sim.png}
\end{figure}

\begin{figure}[htbp]
\centering
\includegraphics[width=\linewidth]{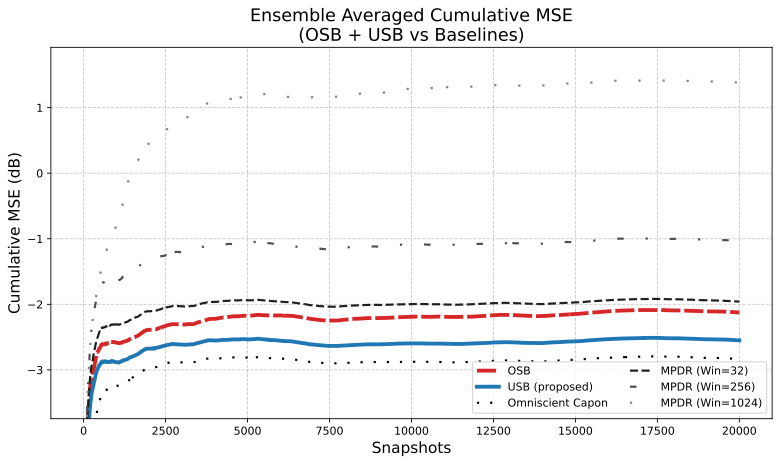}
\caption{Cumulative MSE of the target signal estimate for the birth death scenario averaged over 200 Monte Carlo trials. The switching algorithm tightly tracks the theoretical limit, confirming its regret-minimizing properties.}
\label{fig:birth_death_mse_sim.png}
\end{figure}

\begin{figure}[htbp]
\centering
\includegraphics[width=\linewidth]{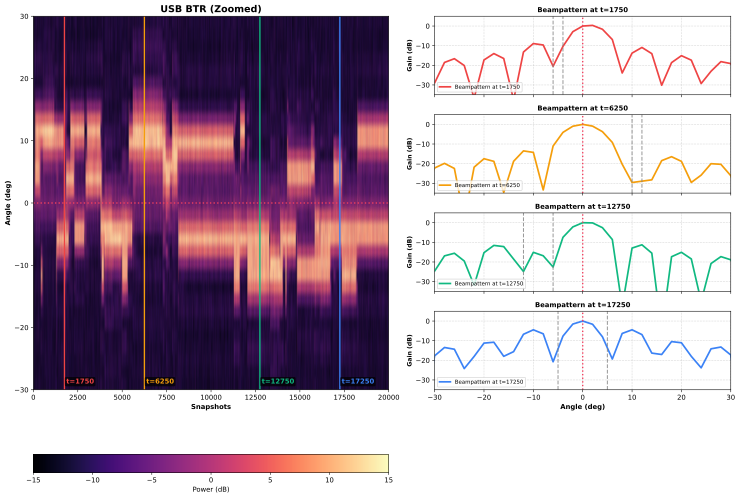}
\caption{Bearing time record and beampatterns at marked times for the universal switching beamformer for the example trial BTR shown for the birth-death simulations.}
\label{fig:pws_birth_death_beampatterns}
\end{figure}

The algorithm's performance is quantitatively validated via output SINR (Figure~\ref{fig:birth_death_sinr_sim.png}) and cumulative MSE (Figure~\ref{fig:birth_death_mse_sim.png}). As in the previous trials, the bearing time record and the beampattern for the universal switching beamformer is shown in Figure \ref{fig:pws_birth_death_beampatterns}.

\section{Experiment - SwellEx96}

\begin{figure}[t]
\centering
\includegraphics[width=\linewidth]{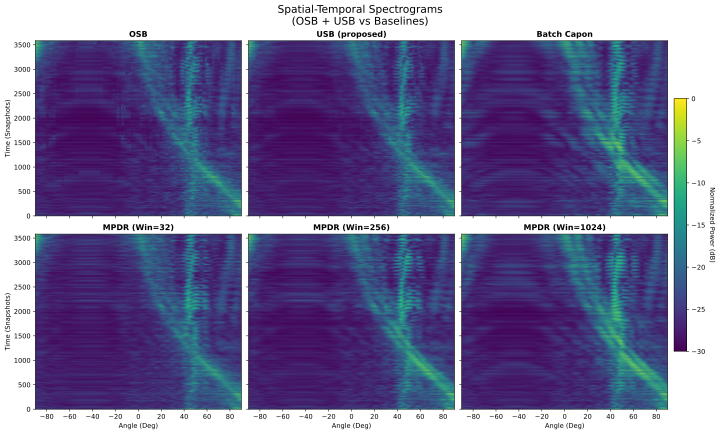} 
\caption{A comparison of different sliding window length beamformers, the sample Capon beamformer, and the USB on the SwellEx-96 S59 data recorded with the Horizontal Linear Array South (HLA-S). CBF refers to the Conventional Beamformer applied without adaptation. All outputs are peak normalized and then plotted with the same dynamic range to enable a fair comparison of the contrast. The USB maintains better noise suppression than short windows and better angular resolution than long windows with moving sources. }
\label{fig:swellex_scanned_response}
\end{figure}

The framework is validated in a complex, real-world environment using the S59 event from the SwellEx-96 experiment. The scanned response is computed in the horizontal plane ($0^\circ$ elevation) to track the source bearing over time. Figure~\ref{fig:swellex_scanned_response} presents the resulting Bearing-Time Records (BTR). 

\begin{figure}[htbp]
    \centering
    \includegraphics[width=\linewidth]{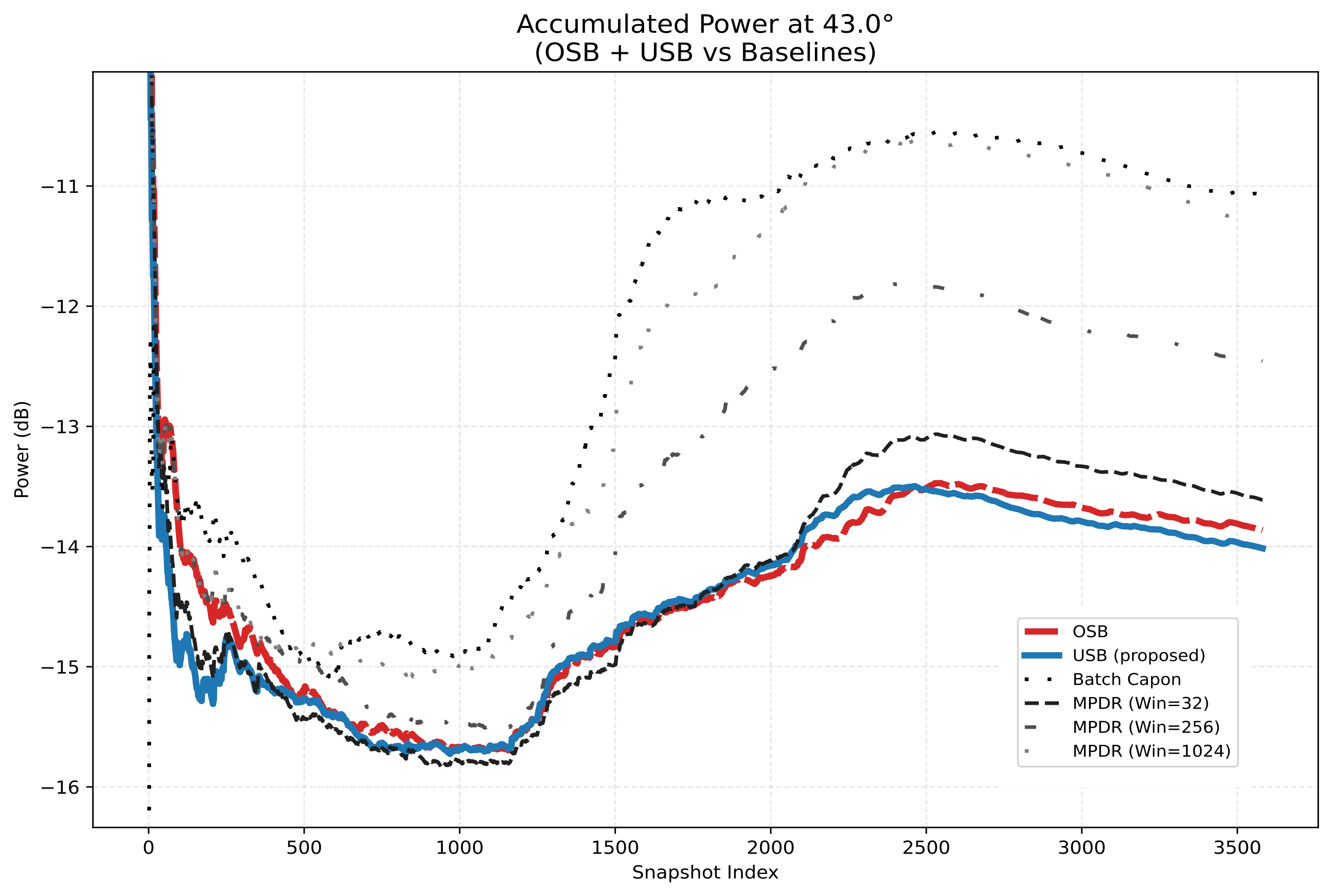}
    \caption{Accumulated output power at $43^\circ$. The USB minimizes its regret over time, performing better with the best sliding window MPDR and the OSB.}
    \label{fig:SwellEx power}
\end{figure}

\begin{figure}[htbp]
    \centering
    \includegraphics[width=\linewidth]{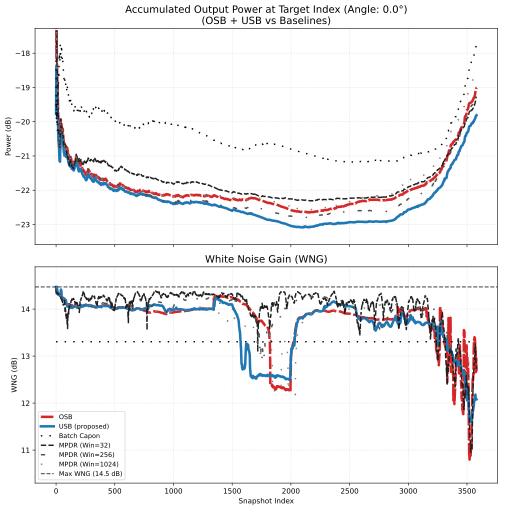}
    \caption{Accumulated output power and white noise gain at $0^\circ$. The USB performs reliably across the target metric dimensions.}
    \label{fig:SwellEx WNG}
\end{figure}

\begin{figure}[htbp]
\centering
\includegraphics[width=\linewidth]{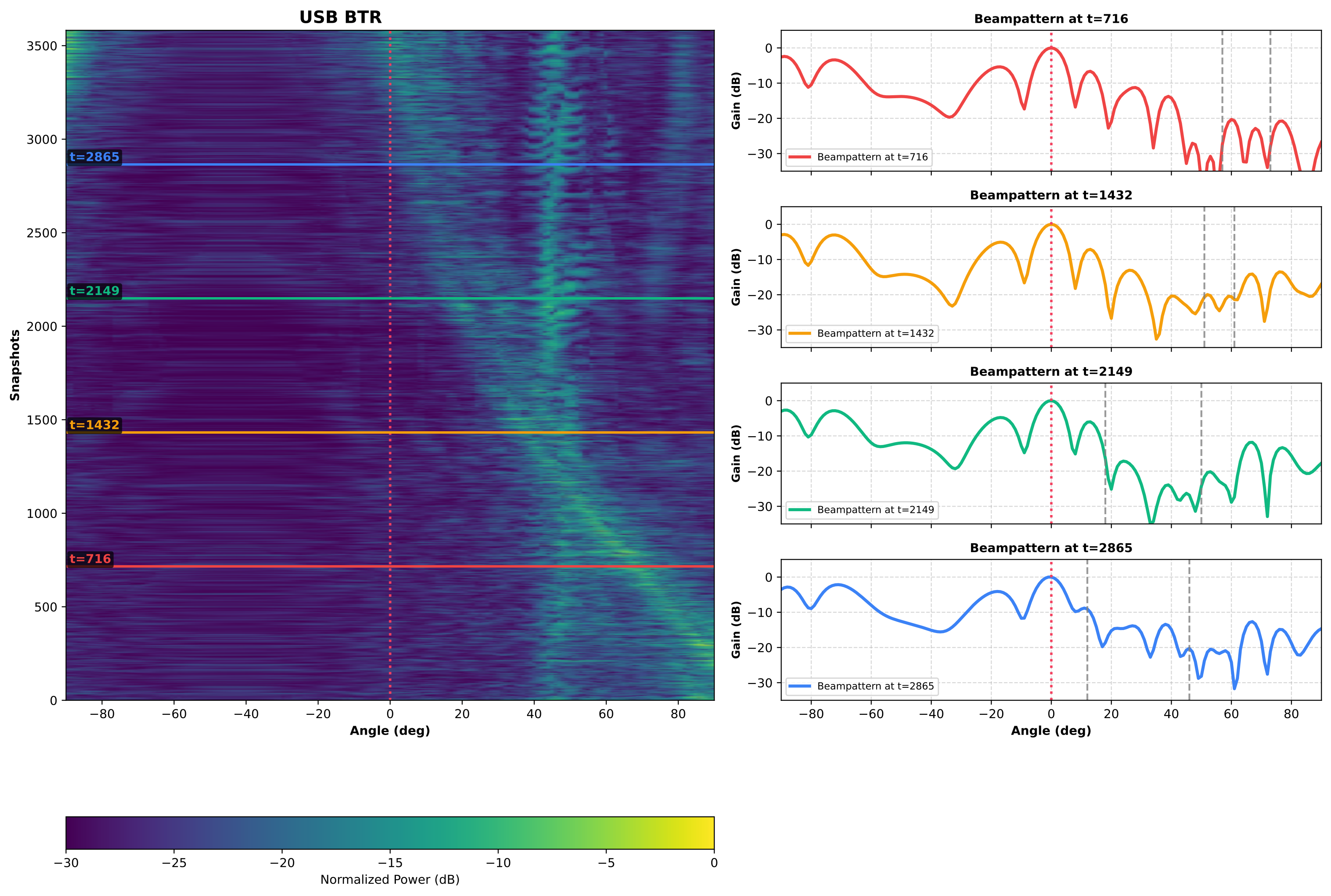}
\caption{Bearing time record and beampatterns for the SwellEx dataset at marked times for the universal switching beamformer steered toward broadside. The plots confirm that the switching beamformer dynamically adjusts the mixture weights to place nulls in relevant locations. Marked interference may contain small estimation errors as ground truth position data is unavailable.}
\label{fig:swellex_beampatterns}
\end{figure}

Performance quantification relies on evaluating the cumulative output power of the beamformers over the observation period. Figure \ref{fig:SwellEx power} shows the accumulated power at $43^\circ$. The USB outperforms the best sliding window beamformer. Figure \ref{fig:SwellEx WNG} compares the white noise gain of the different beamformers and the accumulated output power at $0^\circ$. Finally, the bearing time record alongside the the beampattern at uniformly selected times is shown in Figure \ref{fig:swellex_beampatterns}. 

\section{Conclusion}
Non-stationary environments force a fundamental trade-off: adaptive systems must track rapid environmental changes while maintaining statistical stability. To overcome this challenge, the universal switching beamformer is introduced (USB). By framing the problem of estimating covariance through the lens of sequential decision theory, the algorithm implicitly maintains and updates a large ensemble of candidate memory lengths. 

Theoretically, this approach satisfies a logarithmic regret bound relative to the best piecewise-stationary beamformer selected in hindsight. Empirically, the switching beamformer consistently outperforms fixed-window MPDR formulations in both plane-wave and reverberant environments. The USB successfully arbitrates between tracking abruptly transitioning environments and reducing variance during stable periods. 

\section*{Acknowledgment}
This work was supported by the US Navy, Office of Naval Research, under award N00014-23-1-2133.

\section*{Author Declarations}
The authors have no conflicts of interest to disclose.

\section*{Data Availability}
The data that support the findings of this study are available from the corresponding author upon reasonable request. The majority of the data employed in this work were either generated using standard physics-based simulations or are available through public repositories.

\bibliographystyle{IEEEtran}
\bibliography{references}
\end{document}